\documentclass[prl,twocolumn,preprintnumbers,letterpaper]{revtex4}
\usepackage{amssymb}

\usepackage{graphicx}
\usepackage{amsmath}
\usepackage{times}
\usepackage{color}

\begin{document}

\title{Universal Five- and Six-Body Droplets Tied to an Efimov Trimer}
\author{Javier von Stecher}
\affiliation{ JILA, University of Colorado and National Institute of Standards and Technology, Boulder, CO 80309-0440}

\begin{abstract}

We explore the properties of weakly bound bosonic states in the strongly interacting regime. Combining a correlated-Gaussian (CG) basis set expansion with a complex scaling method, we extract the energies and structural properties of bosonic cluster states with $N\le6$ for different two-body potentials. The identification of five- and six-body resonances attached to an excited Efimov trimer provides strong support to the premise of Efimov universality in bosonic systems.
 Our study also reveals a rich structure of bosonic cluster states. Besides the lowest cluster states which behave as bosonic droplets, we identify cluster states weakly bound to one or two atoms forming effective cluster-atom ``dimers'' and cluster-atom-atom ``trimers.'' The experimental signatures of these cluster states are discussed.
\end{abstract}

\maketitle



The understanding of the universal nature of low-energy few-body body physics is a fundamental prerequisite for the development of effective many-body  descriptions in the ultracold regime. In two-component Fermi gases, the characterization of two-body physics in terms of a single interaction parameter (the scattering length) is at the heart of our understanding of the BCS-BEC crossover. In bosonic systems, the underlying few-body physics is enriched by Efimov physics which leads to the formation of a series of trimers which acquire peculiar properties such as a borromean nature and a discrete scale invariance~\cite{efim70,braaten2006ufb}.
Recent experimental developments  have opened a new era in the exploration of Efimov phenomena in ultracold gases~\cite{kraemer2006eeq,ferlaino2009evidence,zaccanti2009observation,pollack2009universality}. While this richness and complexity of three-boson systems have led to a new level of understanding of universal few-body physics, they pose important questions as to whether the low-energy behavior of larger bosonic systems can be understood and characterized within a simple universal framework.

The natural starting point for addressing this question is the exploration of Efimov and universal phenomena in increasingly larger systems.   For $N>3$, a quantitative analysis of universality is significantly more challenging because it usually requires the study of resonances rather than bound states, since bound states occur in a regime where the applicability of universality is questionable. However, in the last few years, tremendous progress has been achieved in the understanding of universality at the four-boson level~\cite{hammer2007upf,von2009signatures,schmidt2010renormalization,d2009universal}. In particular, a universal regime has been recognized in which the four-body physics is determined by the underlying two- and three-body physics. Recently, some of these early four-body predictions have been quantitatively verified with remarkable accuracy~\cite{deltuva2010efimov}. Despite these important advances, there still remain controversies regarding the scope of universality, the role of the four-body parameter, and the nonuniversal corrections~\cite{thogersen2008n,yamashita2010universality,hadizadeh2011universality}. For $N>4$ systems, the applicability of universal theory is even more debatable because of the lack of theoretical or experimental evidence of universal behavior. A natural continuation to the four-body predictions of Ref.~\cite{von2009signatures,hammer2007upf} would indicate that universality extends to larger clusters, whose behavior follows the same Efimov discrete scale invariance and is only controlled by two and three-body physics. Thus, for each Efimov state ($N=3$), there would be a series of $N$-body states (or resonances) with $N>3$ associated to it forming an {\it Efimov family}. A recent study~\cite{stecher2010weakly}, based on such premise of universality, has characterized some of properties of the lowest weakly bound cluster states up to $N\le13$.
While this study provides key predictions to be theoretically and experimentally explored, it leaves open questions such as
the validity of
 the universality hypothesis and 
 the existence of additional weakly bound cluster states.

In this article, we address these questions through the analysis of the structure of the strongly interacting few-boson spectra.
First, we explore the universal regime and we identify five- and six-boson resonances tied to an excited-Efimov state.
 These resonant states represent small bosonic droplets and are in good qualitative agreement with the prediction of Ref.~\cite{stecher2010weakly}. Our results provide much-needed support to the premise of universality and discrete scale invariance in few-boson systems.
We extend our analysis to the exploration of different types of cluster states formed in the lowest Efimov family, which is strongly modified by nonuniversal corrections. For a range of model potentials, we find that cluster states ($N=3,4,5$) are likely to bind weakly to atoms forming  effective cluster-atom ``dimers'' and cluster-atom-atom ``trimers.''
We identify one of these cluster-atom-atom ``trimers'' as a resonance that appears energetically slightly below the lowest Efimov trimer and it can be qualitatively described as an Efimov trimer formed by an Efimov trimer and two atoms. This state is one of the simplest bizarre cluster structures mathematically proposed~\cite{baas2010new}; similar cluster-atom-atom structures are expected for larger systems.  Our studies reveal an intricate structure of  bosonic cluster states and provide estimates of the $N$-body resonant positions relevant to experiments.

Our starting point is the few-boson Hamiltonian,
\begin{eqnarray}
\label{eq_ham}
\mathcal{H} = -\sum_{i} \frac{\hbar^2}{2m}\nabla_i^2+\sum_{i<j} V(r_{ij}),
\end{eqnarray}
where $m$ is the mass of the bosons and $r_{ij}$ is the interparticle distance between particles $i$ and $j$. The two-body model potential takes form $V(r)=V_0(\exp[-r^2/(2d_0^2)]-\alpha\exp[-2r^2/d_0^2])$, where $d_0$ is the interaction range, and $V_0$ and $\alpha$ are tuned to change the shape and the scattering length of the potential.
We consider two qualitatively different potential: purely attractive ($V_0<0$ and $\alpha\le1$) and attractive with repulsive core potentials ($V_0<0$ and $\alpha > 1$).
 While the studies considered a $-20<\alpha<20$, most of the results will be presented for two cases: a purely attractive interaction $V_a$ ($\alpha=0$) and an attractive potential with a soft repulsive core $V_r$ ($\alpha=2$ and $V_0<0$).  The range $r_0$ and the energy $E_{sr}\equiv\hbar^2/(m r_0^2)$ are the typical energy and length scales that characterize the interactions. The universality regime is characterized by energies $|E|\ll E_{sr}$ and length scales (characterizing scattering length and clusters sizes) $\ell\gg r_0$.



To describe the bound states, we use a CG basis set expansion~\cite{suzuki1998sva,stech07} that has been very successful in describing bosonic and fermionic systems with short-range interactions. In our implementation, the eigenstates of a system are expanded in the set of CG basis functions in which the center-of-mass coordinate has been removed, the relative angular momentum is zero and the parity is positive ($L^\pi=0^+$).  Each basis function is a symmetrized product of Gaussian functions, each of which depends on one of the $N(N-1)/2$ interparticle distances and can be written as $\psi_\beta(\mathbf{x})=\exp\left(-\sum_{ij} A^{\beta}_{i,j}\mathbf{x}_i\cdot\mathbf{x}_j/2\right)$,
where $\mathbf{x}=\{\mathbf{x}_1,\mathbf{x}_2,...,\mathbf{x}_{N-1}\}$ is a set of (relative) Jacobi coordinates.
The parameters $A^{\beta}_{i,j}$ that characterize the Gaussian function widths are selected and optimized using a stochastical variational method~\cite{suzuki1998sva}. The convergence of the results is carefully analyzed by increasing and reoptimizing the basis set. Typical calculations for $N>3$ include up to 3500 optimized and fully symmetrized basis function.
To explore the structure of the few-body states, we extract the pair-distribution function defined as $4\pi r^2P_N(r)=\langle\Psi_N|\delta(r_{ij}-r)|\Psi_N\rangle$, where $r_{ij}$ is the interparticle distance between particles $i$ and $j$.

%
%
%
%
%
%
%
%
%

To study resonances, we use the complex-scaling method (CSM)~\cite{ho1983method,usukura2002resonances,MezeiVargaFBS06}.
In the CSM, all coordinates are rotated as $\mathbf{r}\to\mathbf{r}e^{\imath \theta }$ by a transformation $U(\theta)$. The wave function of the resonance is square-integrable in these rotated coordinates and can be expanded in the same square integrable basis functions that describe bound states:
\begin{equation}
\Psi_\theta(\mathbf{x})\equiv U(\theta) \Psi(\mathbf{x})=\sum_i C_i(\theta) \psi_i(\mathbf{x}),
\end{equation}
where $\mathbf{x}=\{\mathbf{x}_1,\mathbf{x}_2,...,\mathbf{x}_{N-1}\}$ is a set of (relative) Jacobi coordinates.
The wave function $\Psi_\theta(\mathbf{x})$ is a solution of the rotated Hamiltonian $\mathcal{H}_\theta=U(\theta)\mathcal{H}U(\theta)^{-1}$ with complex energy $E_\theta=E_R-\imath \Gamma/2$, where $\Gamma$ is associated with the width of the resonance.

\begin{figure}[h]
\begin{center}
\includegraphics[scale=0.6,angle=0]{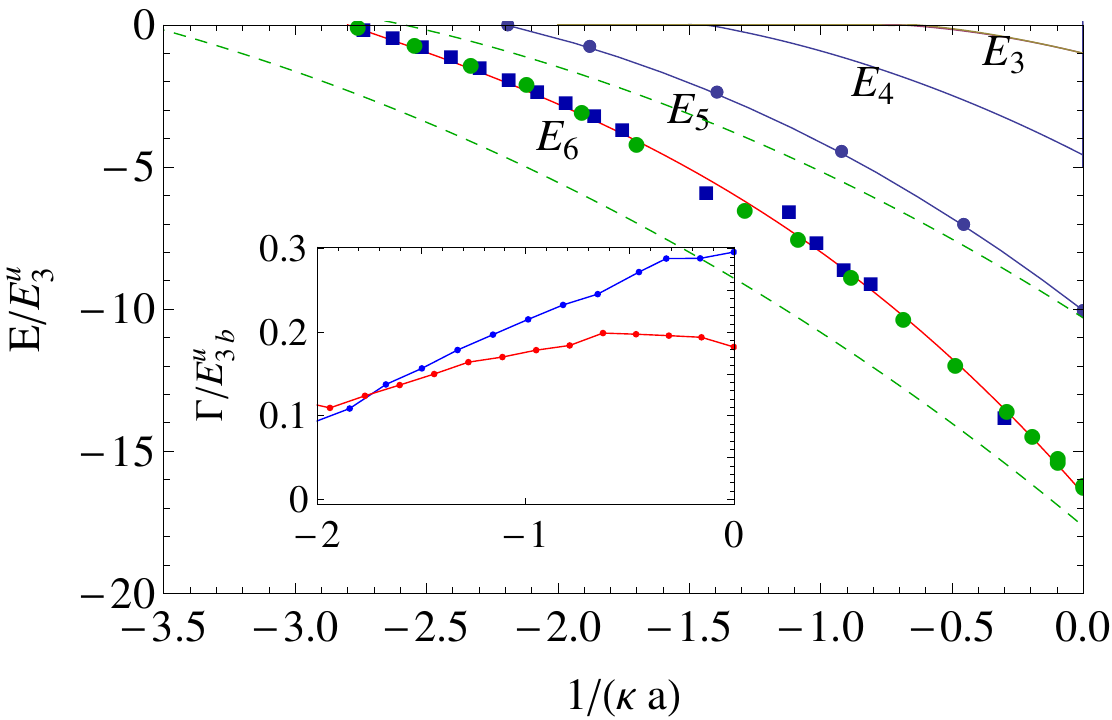}
\caption{(color online) Universal cluster energies as a function of inverse scattering length. The energies for $N=3$ and $N=4$ were obtained in previous studies~\cite{von2009signatures}. The $N=5$ solid curve and symbols correspond to predictions from an excited resonance for $V_r$ and $V_a$ respectively. The circles and squares  are the $N=6$ predictions for $V_r$ and $V_a$ respectively and the solid line is a guide to the eye. Dashed curves correspond to the $N=5,6$ predictions from Ref.~\cite{stecher2010weakly}. Inset: Width of the five-body resonances for $V_r$ (red) and $V_a$ (blue). }\label{SpectUni}
\end{center}
\end{figure}

\begin{figure}[hp]
\begin{center}
\includegraphics[scale=0.55,angle=0]{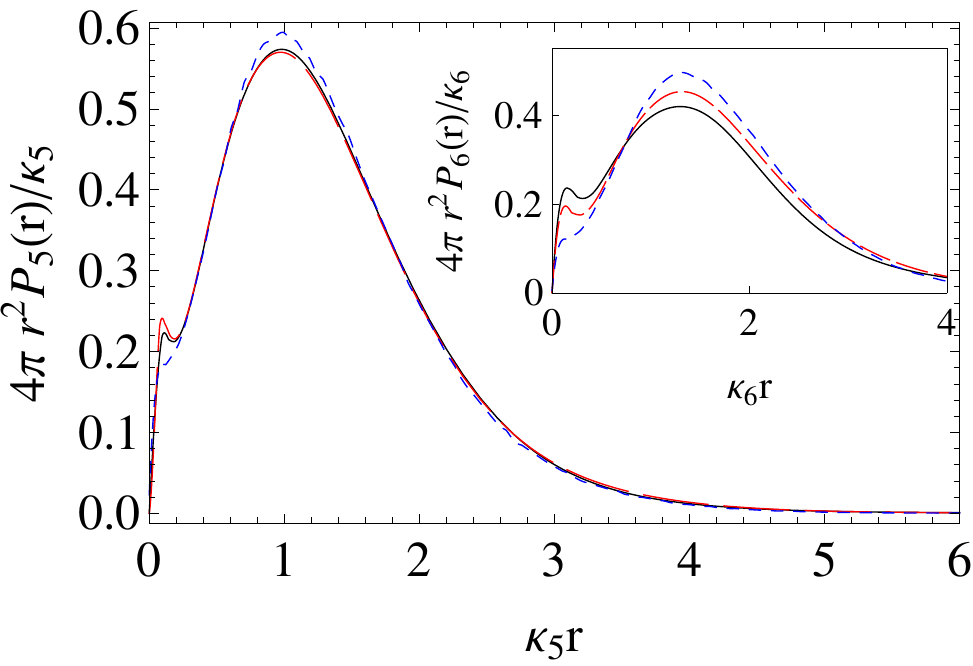}
\caption{(color online) Pair distribution function of the universal five-body state. Here $\kappa_N=\sqrt{m E^u_N/\hbar^2}$, where $E^u_N$ is the binding energy of the $N$-body droplet. Solid and long-dashed curves are the prediction from the $V_a$ and $V_r$, potential, while the short-dashed curve is the prediction from Ref.~\cite{stecher2010weakly}. Inset: pair distribution function for the six-body state. Same curve style convention as the main figure.  }\label{pcuniv}
\end{center}
\end{figure}

{\it Universal Droplets.}--
Figure~\ref{SpectUni} summarizes the energies of the universal cluster states for $N=3-6$. The results are presented as a function of the relevant universal parameters: $1/\kappa a $ and $E/E^u_3$, where $\kappa=\sqrt{m E^u_3/\hbar^2}$ is the three-body parameter, and $E^u_3$ is the binding energy of the Efimov trimer at unitarity ($a=\infty$). The five- and six-body results are obtained from an analysis of resonances attached to an excited Efimov family.
These resonances are observed for both the $V_a$ and $V_r$ potentials. At unitarity, we obtain $E_{5}^u\approx 10.1 E^u_3$ in close agreement of Ref.~\cite{stecher2010weakly} predictions of $E_{5}^u\approx 10.4(2) E^u_3$. For $N=6$, we find that $E_{6}^u\approx 16.3 E^u_3$ which is slightly lower than the predictions from Ref.~\cite{stecher2010weakly}  of $E_{6}^u\approx 18.4(2) E^u_3$. Interestingly, the energies of the universal states at unitarity scales roughly with the number of trimer configurations the cluster supports, i.e., 1, 4, 10, 20 for $N=3$, 4, 5, and 6.
 To further verify the agreement between different model potential-predictions, we analyze the pair distribution function at unitarity (see Fig.~\ref{pcuniv}). The good agreement between three different predictions of $P_5$ and $P_6$ demonstrate the universality of such few-body resonances.

The widths of the four- and five-body resonances depend strongly on the open decay channels. In our calculations, resonances of the first excited Efimov family can only decay into the lowest Efimov family, i.e. into decay  channels that are not quantitative in the universal regime.
 Therefore we expect the width of the resonance to be more sensible to non-universal corrections. For example, our analysis of the first-excited four-body resonance leads to a $\Gamma^u_{4b}\approx 0.1 E^u_3$, which is a factor of 3 larger than the predicted-converged width Ref.~\cite{deltuva2010efimov} (similar deviations were reported~\cite{deltuva2010efimov}). The width of the first excited five-body resonance is presented in Fig.~\ref{SpectUni} for both $V_a$ and $V_r$. The clear differences between the different $\Gamma$ predictions illustrate the importance of the nonuniversal corrections for resonances belonging to the lowest-excited Efimov families. However, as in the four-body case, we expect that the width extracted from the first-excited five-body resonance provides a correct order of magnitude estimate of the widths in the universal limit. We also estimate $\Gamma^u_{6b}\sim 0.3 E^u_3$.



%
%

%
%

The description of these five- and six-body resonances is extremely challenging since it entails an exploration of energies that are  a factor of $e^{2\pi/s_0}\approx 515$ smaller than the energies of the lowest fragmentation threshold.
To obtain an accurate representation of these resonances, we carry out a numerical procedure inspired by previous implementations of the SVM+CSM~\cite{usukura2002resonances,MezeiVargaFBS06}.
First, we generate a basis optimized to describe a wide range of energies.  We introduce an external trapping potential $V_{ext}(\mathbf{r})=m \omega^2 \mathbf{r}^2/2$  in this part of the optimization to reduce the number of states described at each fragmentation threshold. As the basis is increased, the trapping frequency is reduced and finally set to zero to extract bound states and resonances. 
The energy trajectories as a function of the basis set size show clear plateaus around the energy resonance that helps us identify it providing first estimates of the real part of the energy.
To improve the description of the resonances, a second optimization procedure is carried out involving only states in an energy window around the resonance. Once the optimization is complete, we calculate the spectrum as a function of $\theta$ to identify the energy and the width of the resonance. The large number of avoided crossings between the resonant state and other states makes it particularly challenging to quantitatively estimate the six-body energy (see symbols in Fig.~\ref{SpectUni}). Taking into account the numerical uncertainties and the potential nonuniversal corrections in the first excited Efimov family, we estimate a 10\% uncertainty in the energies of resonances and the positions of the resonances discussed below.



%
\begin{figure}[h]
\begin{center}
\includegraphics[scale=0.55,angle=0]{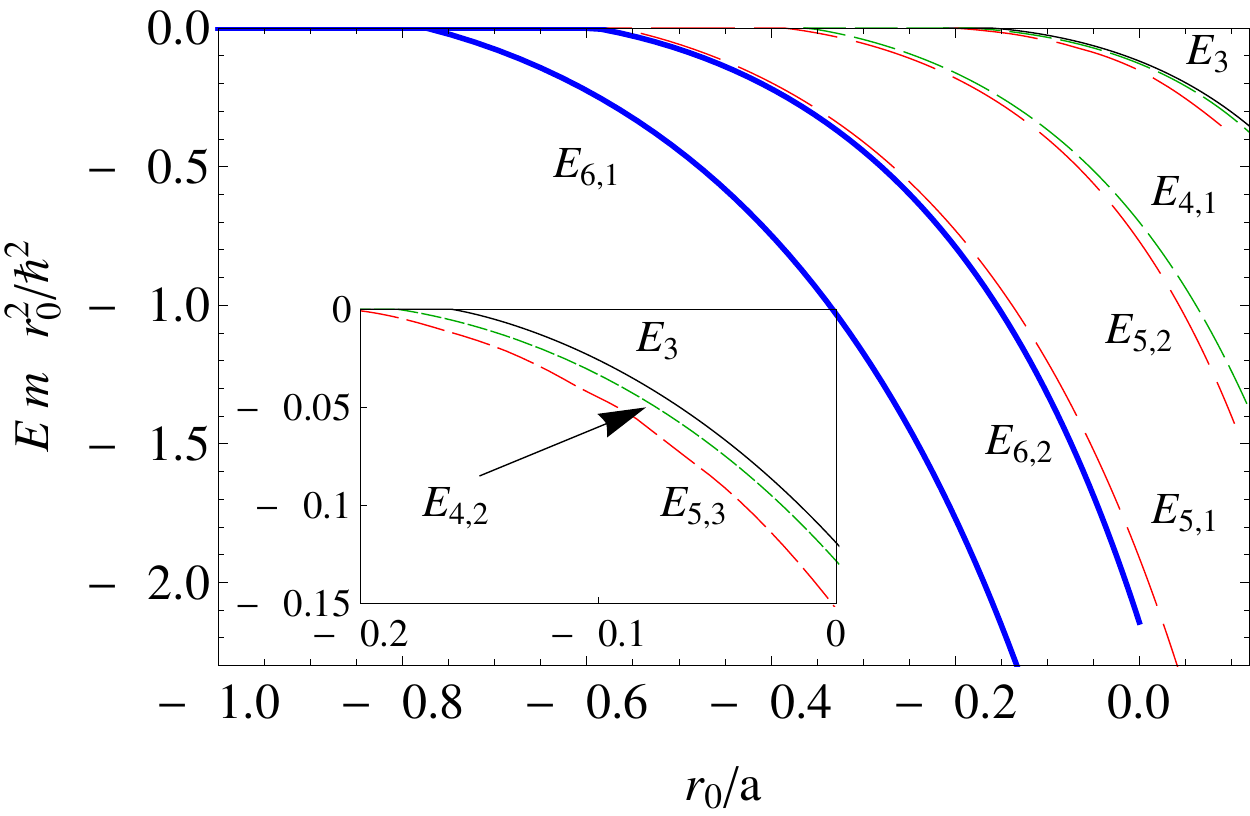}
\caption{(color online) Spectrum as a function scattering length for $3\le N\le6$. The thin solid curve corresponds to the trimer state energy, the short dashed curves correspond to the four-body states, the long-dashed curves correspond to the five-body states, and the thick solid curves correspond to the six-body states. Inset: States formed below the trimer threshold.}
\label{ClusSpect}
\end{center}
\end{figure}

{\it Excited cluster states and nonuniversal corrections.}--
Next, we analyze the formation of bound states for $3<N\le6$ in the lowest Efimov family. This study illustrates the structure of the spectrum in the strongly interacting regime by identifying other types of bound states. 
It also addresses the issue of nonuniversal corrections
in the regime $|a|\gtrsim r_0$ that are particularly important for understanding  $^{133}$Cs and $^7$Li Efimov experiments in which five- and six-body resonant phenomena are expected to occur at  $|a|\lesssim 3 r_{vdw}$, where $r_{vdw}$ is the Van der Waals length.
The general structure of the bosonic spectrum is shown in  Fig.~\ref{ClusSpect}. These results correspond to the potential $V_a$. This structure, although it changes quantitatively, remains qualitatively the same for a range of model potentials.
The lowest $N$-body state 
is analogous to the universal states of Fig.~\ref{SpectUni}. However, the energy of the lowest $N$-body states grows very fast with the number of particles, implying that nonuniversal corrections increase with $N$ (in agreement with Ref.~\cite{yamashita2010universality}). For example, the energy per particle of the lowest trimer state of $V_a$  at unitarity is $\sim 0.04  E_{sr}$, while the energy per particle of the lowest six-body state at unitarity is $\sim 0.6E_{sr}$. The latter result implies that $E^u_6/E^u_3\sim 30$, almost a factor of two larger than the universal predictions. The introduction of a repulsive three-body force, as proposed in Ref.~\cite{stecher2010weakly}, leads to a ratio of $E^u_6/E^u_3\sim 18$, which is significantly closer to the universal predictions.

\begin{figure}[h]
\begin{center}
\includegraphics[scale=0.5,angle=0]{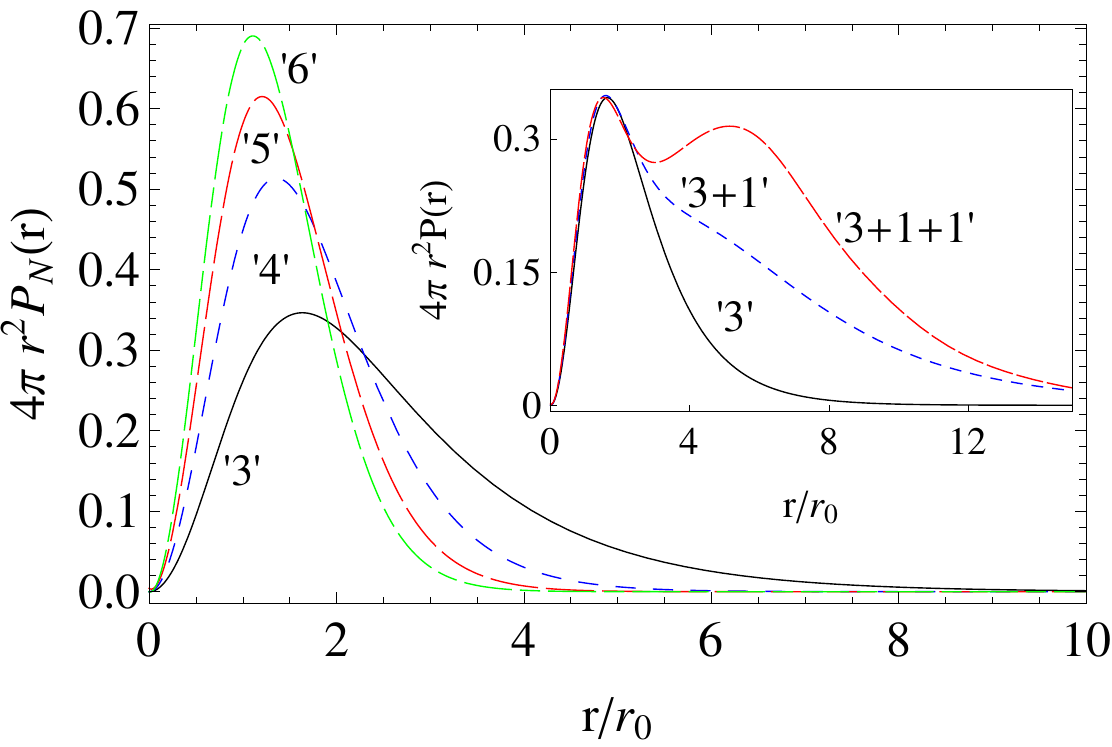}
\caption{(color online) Pair distribution functions of the lowest $N$-body cluster states ($N=3-6$). Inset: Distribution functions for the trimer ('3'), the trimer-atom four-body state ('3+1'), and the trimer-atom-atom five-body state ('3+1+1').  }\label{PCLowest}
\end{center}
\end{figure}

The increasing importance of nonuniversal corrections as $N$ increases is also reflected in the pair distribution functions presented in Fig.~\ref{PCLowest}. As $N$ increases, the lowest cluster states become more localized in the nonuniversal region ($r\lesssim r_0$) and, therefore, become less universal. The single peak structure of the pair distribution indicates that these states are basically droplets which are mainly described by configurations at which all particles are   at similar distances.

The excited states (with energies $E_{4,2}$, $E_{5,2}$, and $E_{6,2}$) are much closer to the lowest fragmentation threshold and can be qualitatively described as cluster-atom ``dimers''  with one particle loosely bound to an $N-1$ cluster state. This structure can also be identified in the pair correlations that coincide at small $r$ with the $N-1$ cluster pair correlation but it has a longer tail that describes the cluster-atom correlation (cf. inset in Fig.~\ref{PCLowest}).

We also identify another five-body resonance, presented in Fig.~\ref{ClusSpect} as $E_{5,3}$, that is a state energetically below the lowest trimer-fragmentation threshold that can decay into the lowest tetramer-atom channel. The energy and the pair-distribution function show that the state is qualitatively described as a trimer weakly bound to two atoms forming a trimer-atom-atom state. The energies at unitarity of such states are  $1.3 E^u_3$ ($V_a$) or $1.2 E^u_3$ ($V_t$), i.e., slightly below the trimer and the excited tetramer energies, suggesting that most of the contribution of the energy comes from the bonding of the trimer subcluster. Furthermore, the pair-distribution function shows two clear peaks that can be identified as coming from atom-atom correlations inside the trimer  subcluster and atom-atom correlations between an atom inside the trimer and an atom outside the trimer. At small $r$, the five-body $P_{5}$ agrees well with the trimer $P_{3}$, suggesting that the small $r$ part of the $P_{5}$ comes exclusively from the trimer subcluster contribution.
The relative strength between the pair-correlation peaks is consistent with this interpretation.

These states can be experimentally observed in ultracold gases through the analysis of $N$-body recombination processes~\cite{mehta2009general}. Current experiments with Cs and Li have identify three- and four-body resonances through the observation of losses at the predicted resonance positions~\cite{kraemer2006eeq,ferlaino2009evidence,zaccanti2009observation,pollack2009universality}.  At low temperatures, the $N$-body resonant enhancement of losses occurs at the critical interaction strengths at which an $N$-body cluster becomes resonant with the free particle-scattering continuum. If the $N$-body clusters behave universally, the positions of the resonances are given by critical scattering lengths $a^*_{N}$, which are only controlled by the three-body parameter; the ratio between any two $a^*_{N}$ is a universal number. We explore the critical scattering-length ratios for the lowest cluster states in a range of model interactions and find that  $0.45\lesssim a^*_{4}/a^*_{3}\lesssim 0.47$ and $0.63\lesssim a^*_{5}/a^*_{4}\lesssim 0.67$. The description of the six-body state for strong repulsive core potentials is challenging and does not allow an accurate estimation of $a^*_{6}$. In the range of potentials where convergence is achieved we find that $a^*_{6}/a^*_{5}\sim 0.73-0.74$.  The four-body predictions are relatively close to the universal prediction of $a^*_{4}/a^*_{3}\approx 0.43$ and its deviations are comparable to those observed in experiments (which also analyze the first Efimov family).
The five- and six-body scattering-length ratios are in the same ballpark of Ref.~\cite{hanna2006energetics} predictions ($a^*_{5}/a^*_{4}\approx 0.69$ and $a^*_{6}/a^*_{5}\approx 0.78$) and of Ref.~\cite{stecher2010weakly} ($a^*_{5}/a^*_{4}\approx 0.6$ and $a^*_{6}/a^*_{5}\approx 0.7$) based on  different model interactions. From the analysis of  the five and six-body resonances we predict universal scattering-length ratio of $a^*_{5}/a^*_{4}\approx0.66$ and $a^*_{6}/a^*_{5}\approx0.78$.

In conclusion, we have demonstrated the existence of universal five- and six-body resonances.  In experiments, these five- and six-body states should manifest as a loss peak at the critical scattering length that should occur at approximately $a^*_{5b}\sim 270 a_0$ and at  $a^*_{6b}\sim 210 a_0$ for the Cs experiments at Innsbruck~\cite{kraemer2006eeq,ferlaino2009evidence}. Our results strongly support the hypothesis of a universal regime in which bosonic systems are only controlled by two- and three-body physics.
 We extend the analysis to the lowest Efimov family and we identify a rich structure of bound states and resonances in bosonic systems with large scattering lengths. This structure can be interpreted as emerging from a universal behavior that is strongly affected by nonuniversal corrections. Therefore, much of this structure is expected to persist in the universal regime. In particular, we expect the persistence of universal trimer-atom-atom resonant states. A recent study~\cite{deltuva2010efimov} predicts a large and positive atom-trimer scattering length, which implies that, even in the universal regime, the trimer-atom-atom system fulfils Efimov conditions for the formation of weakly bound trimers.
The emergence of universal picture for bosons suggests a reinterpretation of previous studies on strongly interacting systems. For example, one can speculate that the similarities found in the formation of $^4$He and Tritium clusters~\cite{blume2002formation} have their root in an underlying universal behavior that has Efimov physics at its root, but is modified by nonuniversal corrections.








Enlightening discussions with N. Baas, M. Berninger, D. Blume, J. P. D'Incao, F. Ferlaino, C. H. Greene, and D. A. Huse are acknowledged.
This work was initiated at the INT-10-46W workshop and partially supported by NSF.


%

\end{document}